\documentclass[a4paper]{aa}
\usepackage{psfig,times}
\begin{document}
%\psdraft
\def\apj{{ApJ}}       
\def\aj{{AJ}}       
\def\apjs{{ Ap. J. Suppl.}} 
\def\apjl{{ Ap. J. Letters}} 
\def\pasp{{ Pub. A.S.P.}} 
\def\mn{{MNRAS}} 
\def\aa{{A\&A}} 
\def\aasup{{ Astr. Ap. Suppl.}} 
\def\baas{{ Bull. A.A.S.}} 
\def\csss{{Cool Stars, Stellar Systems, and the Sun}\ }
\def\an{{Astron. Nachr.}}
\def\sp{{Solar Phys.}}
\def\gafd{{Geophys. Astrophys. Fluid Dyn.}}
\def\ass{{Ap\&SS}}
\def\acta{{Acta Astron.}}
\def\jfm{{J. Fluid Mech.}}
\def\ea{{et\thinspace al.\ }}   
\def\AIP{Astrophysikalisches Institut Potsdam}
\def\gR{G. R\"udiger}
\def\lK{L.L. Kitchatinov}
\def\mK{{M.~K\"uker}\ }
\def\S{St\c{e}pie\`{n}}
\def\DR{differential rotation\ }

\def\Oms{ {\Omega^*} }
\def\O2{ {\delta \tilde{\Omega} }}
\def\bib{\item}
\def\beg{\begin{equation}}
\def\ende{\end{equation}}

\title{Meridional flow and differential rotation by 
gravity darkening in  fast rotating  solar-type stars }
\author{G. R\"udiger \inst{1}
        \and\  M. K\"uker\inst{1,2}}
\offprints{G. R\"udiger}
\institute{Astrophysikalisches Institut Potsdam,
An der Sternwarte 16, D-14482 Potsdam, Germany \and   
Astrophysikalisches Institut
        und Universit\"atssternwarte,
        Schillerg\"asschen 2-3, 07745 Jena, Germany}
\date{Received date; accepted date}
\abstract{An explanation is presented for the  rather strong total surface 
differential rotation of the observed very young solar-type stars like AB Dor and 
PZ Tel. Due to its rapid 
rotation a nonuniform energy flux leaves the stellar core so that the 
outer convection zone is nonuniformly heated from below. Due to this `gravity 
darkening' of the equator a meridional flow is created flowing equatorwards at 
the surface and thus accelerating the equatorial  rotation.  
The effect linearly grows with the normalized pole-equator difference,
$\epsilon$, of the heat-flux at the bottom of the convection zone. A rotation
rate of about 9 h leads to $\epsilon=0.1$ for a solar-type star. In this case the resulting 
equator-pole differences of the angular velocity at the stellar surface,
$\delta\Omega$, varies from  unobservable 0.005 day$^{-1}$ to the (desired) value of  0.03 day$^{-1}$ when the dimensionless diffusivity factors $c_\nu$
and $c_\chi$ vary between   1 and 0.1 (standard value $c_\nu \simeq c_\chi \simeq
0.3$, see Table \ref{tab1}.) In all cases  
the related temperature differences between pole and equator at the surface are unobservably small. \\
The  (clockwise) meridional circulation which we obtain flows  opposite to  the (counterclockwise) circulation  appearing   as a byproduct in the  $\Lambda$-theory of the nonuniform  rotation in outer convection zones.  The consequences of this situation for those dynamo theories of stellar activity are discussed which
 work with the meridional circulation as the dominant magnetic-advection effect in latitude to produce the solar-like form of the butterfly diagram. 
}

\titlerunning{Differential rotation and gravity darkening for fast rotating stars}
\authorrunning{G. R\"udiger \& M. K\"uker}

\maketitle

%%%%%%%%%%%%%%%%%%%%%%%%%%%%%%%%%%%%%%%%%%%%%%%%%%%%%%%%%%%%
\section{Introduction}
In a series of papers the  internal  rotation  of outer convection zones of cool stars have been derived using  a consistent mixing-length model of Reynolds stress and convective 
heat transport in a rotating convection zone that takes into account the
effect of the Coriolis force on the convective motions   (K\"uker \ea 1993, Kitchatinov \& R\"udiger 1995 (KR95),    Kitchatinov \& R\"udiger 1999 (KR99)  ).
In the KR95 model, the deviation of the heat flux from spherical symmetry
causes a small horizontal temperature gradient, which partly neutralizes
the rotational shear as the force that drives the meridional flow.
As a result, the model reproduces the original K\"uker \ea (1993) result
with only one free parameter, the mixing-length parameter $\alpha_{\rm MLT}$.
\par
KR99  found
the total shear, 
\begin{equation}
 \delta \Omega  =  \Omega_{\rm eq} - \Omega_{\rm Pole},
\label{1}
\end{equation}
to be only rather weakly dependent on the rotation period, $\Omega$.
Collier Cameron \ea (2001) have combined measurements of differential
rotation for the pre-main-sequence stars RX J1508-4423, AB Dor, and PZ Tel
with the sample of main-sequence stars from Donahue \ea (1996)
and argue that the discrepancies between the KR99
result and the relation between shear and rotation
rate derived by Donahue \ea(1996) may be due to the mix of G and K stars 
in the latter paper.
They suggest that $\delta \Omega$ may indeed be constant with rotation
rate but vary with spectral type and derive values for G and K dwarfs
that differ by a factor of three, the K dwarfs rotating more rigidly. His
discussion of the sample of photometric data for several classes of stars has led
Hall (1991) to the surprisingly flat distribution profile $\delta\Omega \propto
\Omega^{0.15}$. Further observations must decide how flat the $\delta \Omega -
\Omega$ relation for solar-type stars really is. 
\par
\begin{figure}
\psfig{figure=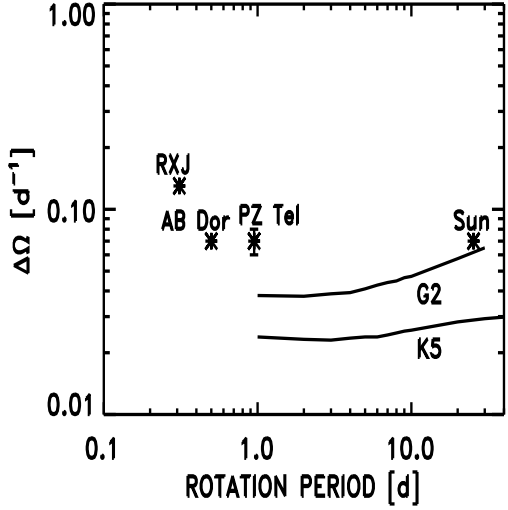,width=7cm,height=7cm}
 \caption{ \label{alpha}
  The total equator-pole difference of the surface angular velocity (1) versus
  the period $\tau_{\rm rot}$ of the basic rotation after
  the model in KR99 and as observed 
 }
\end{figure}

In Fig.~\ref{alpha} the equator-pole differences (\ref{1}) for the known rapid 
rotators are plotted in comparison to the results of the simulations by 
KR99. Obviously, the rapid rotators exhibit more 
differential rotation than the computations predict. One could believe -- and 
we do -- that another reason for differential rotation exists, which becomes 
important only for very rapid rotation.
\par
In the present paper the idea is adopted that for very fast rotation a nonuniform 
energy flux leaves the stellar radiative core (`gravity darkening'), thus heating from 
below the 
stellar convection zone nonuniformly (\S\ \ea 1997). This way,  a meridional 
circulation should arise in the convection zone which even by itself produces a 
nonuniform surface rotation (Kippenhahn 1963).  We shall see that an accelerated equator is the natural consequence with an amplitude explaining indeed the surprisingly steep surface rotation laws which are performed by the fast rotators.

%%%%%%%%%%%%%%%%%%%%%%%%%%%%%%%%%%%%%%%%%%%%%%%%%%%%%%%%%%%%%%
\section{The Model}
In the diffusion approximation the radiative flux in a radiative stellar shell is 
proportionate to the gradient of the temperature which itself is a function of 
the total gravitational potential $\phi = \psi - \int\limits_0^s \Omega^2 s'ds'$. 
Hence
\beg
\vec{F}= f(\phi) \nabla\phi = -f(\phi) \vec{g}^{\rm eff}
\label{2}
\ende
with the effective gravity acceleration $\vec{g}^{\rm eff} = \vec{g} - 
\Omega^2 \vec{s}$. A 
convection zone above the radiative core is thus more intensely heated from
below  at the pole and it is less intensely heated from below at the equator. 
 The intensity of this effect is given by the dimensional parameter 
\beg
\epsilon = {\Omega^2 R\over g}.
\label{3}
\ende
Its value for solar-type stars is $\epsilon\simeq 0.014/(\tau_{\rm rot}
/{\rm day})^2$ with 
$\tau_{\rm rot}$ as the rotation period. For the very rapid rotation of 0.1 days the 
$\epsilon$ is of order unity and the star breaks off. In our computations the 
rotation parameter $\epsilon$ is considered as a free parameter, the models concern 
to a star with a rotation period of 9 h, i.e. $\epsilon \simeq 0.1$. The same value would be  true  for a K giant with 19 days rotation period. The K0 giant KU 
Peg, however,  has only  $\epsilon=$0.04 (see Weber \& Strassmeier, 2001, for  differential rotation and meridional flow\footnote{Equatorial acceleration and 
poleward flow, quite similar as at  the Sun}).
\subsection{The transport equations}

We use the mean-field formulation of hydrodynamics, i.e. apply an appropriate
averaging procedure to split the velocity field into a mean and a fluctuating
part. 
The mean velocity field, $\bar{\vec{u}}$, is then governed by 
the Reynolds equation,
\begin{equation} \label{reynolds}
  \rho \left [ \frac{\partial \vec{\bar{u}}}{\partial t}
      + (\vec{\bar{u}}\cdot \nabla)
       \vec{\bar{u}} \right ] =  - \nabla \cdot (\rho Q)
	   - \nabla p + \rho \vec{g},
\end{equation}
where
$Q_{ij}=\langle u_i' u_j' \rangle$ is the correlation tensor of the
velocity fluctuations, $\vec{u}'$. 
\par
As the density distribution is spherically symmetric, and the gas motion
is dominated by the global rotation, we assume axisymmetry for the mean
velocity and temperature.  The velocity field can then be 
described as a superposition of a rotation and a meridional flow, $\vec{\bar{u}}=r \sin \theta \, \Omega \,
			   \vec{\hat{\phi}} + \vec{\bar{u}}^{\rm m}$,
where $\vec{\hat{\phi}}$ is the unit vector in the azimuthal direction.
The azimuthal component of the Reynolds equation expresses the conservation
of angular momentum:
\begin{equation} \label{omega}
    \frac{\partial \left(\rho r^2 \sin^2 \theta \: \Omega\right)}{\partial t}
     + \nabla \cdot \vec{t} = 0,
\end{equation}
where
$\vec{t} = \rho r \sin \theta \left [r \sin \theta
    \Omega \vec{\bar{u}}^{\rm m}
    + \langle u_{\phi}' \vec{u}' \rangle \right].
$
As the mass density varies with depth but not with time, mass
conservation requires 
$
 \nabla \cdot (\rho \vec{\bar{u}}) = 0.
$
The meridional circulation can then be expressed by a stream function $A$.
The stream function and the vorticity $\omega$ are related via the
equation
\begin{equation} \label{poisson}
 D A - \frac{1}{\rho}\frac{\partial \rho}{\partial r} \frac{\partial A}
	{\partial r} - \frac{1}{\rho r} \frac{\partial \rho}{\partial r} A
	= - \rho \omega,
\label{curl}
\end{equation}
where 
$
 D = \Delta - {1}/(r^2 \sin^2\theta).
$
\par
The correlation tensor $Q$ consists of a viscous and a non-viscous part. 
The eddy viscosity tensor,
\begin{equation}
 Q^\nu_{ij} = - {\cal N}_{ijkl} \frac{\partial \bar{u}_k}{\partial x_l},
\end{equation}
has been calculated by Kitchatinov \ea(1994). It is here simplified to ${\cal N}
= \nu_{\rm T}(\delta_{ik} \delta_{jl} + \delta_{il} \delta_{jk})$ ignoring the
effect of the basic rotation to the tensorial structure. For the eddy viscosity
$\nu_{\rm T}$ the mixing-length expression
\beg
\nu_{\rm T} = c_\nu \tau_{\rm corr} u_{\rm T}^2
\label{nue}
\ende
is used where as usual $\tau_{\rm corr}$ is the convective turnover  time and
$u_{\rm T}^2$ the turbulence intensity. The dimensionless parameter $c_\nu$ is
free but it should not exceed unity. Its standard value is 0.3 but under the
influence of rotation also smaller values are possible.
The non-viscous $\Lambda$-effect
is the main source of differential rotation. In order to select the influence 
of the gravity darkening of the equator it is neglected, however, in the 
following calculations.
The convective heat transport is described by the transport equation
\begin{equation} \label{heat}
\rho T \frac{\partial s}{\partial t} 
                + \rho T \vec{u} \cdot \nabla s  
          = - \nabla \cdot  (\vec{F}^{\rm conv} + \vec{F}^{\rm rad})
           + q,
\end{equation}
where  $q$ is the source function. In case of a perfectly 
adiabatic stratification the entropy is constant throughout the whole
convection zone. 
Standard mixing-length theory does not include the rotational influence
on the turbulent heat transport.
Kitchatinov \ea(1994) have applied the same turbulence model as in their
calculation of the Reynolds stress and found 
\begin{equation}
 \vec{F} = \rho T \chi_{\rm T} \vec{\nabla} s,
\end{equation}
where
\begin{equation}
\chi_{\rm T} = c_\chi \tau_{\rm corr} u_{\rm T}^2
\end{equation}
is  
 the mixing-length expression for the heat conductivity coefficient
 in a slowly rotating convection zone, i.e.~$\Omega^* \ll 1$.
Again $c_\chi$ is a free parameter. More details of the corresponding thermodynamics are given by K\"uker \& Stix 
(2001).
\subsection{Model of the convection zone and boundary conditions}
KR95 used a simplified model of the convection zone which is derived
from the full model  by solving the equations
%\begin{eqnarray*}
$$
\frac{dT}{dr} = -\frac{g(r)}{C_p}, \hspace{5mm}
\frac{dg}{dr} = -2\frac{g}{r}+4 \pi G \rho, \hspace{5mm} 
\rho  =  \rho_e (\frac{T}{T_e})^{\frac{1}{\gamma-1}}
$$
%\end{eqnarray*}
from a starting point $x_e$,
where the reference values $g_e$, $\rho_e $ and $T_e$ are
taken from a standard solar model. Together with the opacity law,
$
  \kappa = 0.34 (1+6\times 10^{24} \rho T^{-7/2}) \hspace{0.2cm} 
         \mbox{\rm cgs},
$
this gives a stratification quite close
to that of the model the reference values were taken from. 
\par
We additionally  require that both the upper and lower boundaries are stress-free,
\begin{equation}
 Q_{r \phi} = Q_{r \theta} = 0.
\end{equation}
As boundary condition for the heat flux  we require that the total 
flux through the outer boundary
is equal to the total luminosity
\begin{equation} \label{frbound}
 F_r^{\rm tot}(r) = \frac{L}{4 \pi r^2}.
\end{equation}
At the inner boundary the condition is now
\beg
F_r^{\rm tot} = {L \over 4\pi r_i^2} \left(1+\frac{\epsilon}{\epsilon+3}
      (3\cos^2\theta-1)\right).
\label{Ftot}
\ende
The parameter $\epsilon$ here gives the normalized difference between the heat
flows at the poles and the equator,
\beg
\epsilon = \frac{F^{\rm tot}_r({\rm pole})
           -F^{\rm tot}_r({\rm eq})}{F^{\rm tot}_r({\rm eq})}.
\ende

We have also tried the outer boundary condition of KR95, 
\begin{equation} \label{leobou}
  F_r^{\rm tot}(r_e) = \frac{L}{4 \pi r_e^2} 
                     (1+4\frac{T \delta s}{C_p T_{\rm eff}})
\end{equation}
and found only small differences. However, (\ref{leobou}) does not 
ensure that the luminosity is constant with radius and the solutions 
show a slow drift in the total entropy while with 
(\ref{frbound}) a stationary state is reached after a few diffusion times.

%%%%%%%%%%%%%%%%%%%%%%%%%%%%%%%%%%%%%%%%%%%%%%%%%%%%%%%%%%%%%
\section{Results}
\begin{figure}
 \hspace{-0.5cm}
\psfig{figure=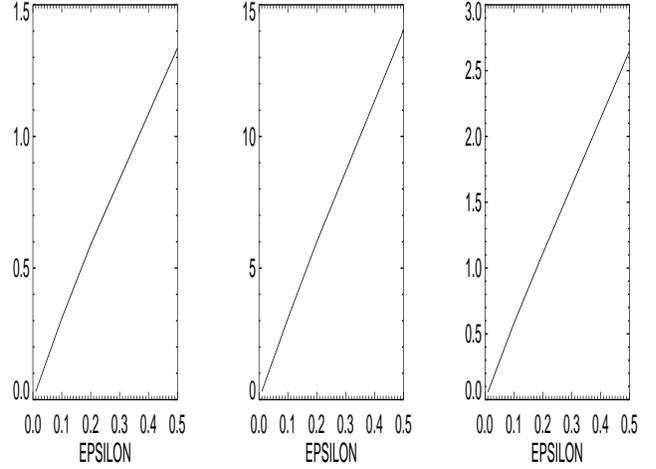,width=9cm,height=7cm}
 \caption{ \label{surf}
  LEFT: The difference between the rotation rates at the equator and the
       the poles in deg/day.
  MIDDLE: The maximum value of the meridional flow velocity at the surface
       in m/s. 
  RIGHT:  The maximum value of the meridional flow velocity at the bottom
       in m/s. 
 $c_\nu=c_\chi=1$
 }
\end{figure}
The results for  $c_\nu=c_\chi=1$ are given in Fig. \ref{surf}. We observe a
nearly linear relation of the total equator-pole difference of the angular
velocity and the amplitudes of the latitudinal drift at the bottom and the 
top of the convection zone with the darkening parameter $\epsilon$. We find
\beg
u_{\rm top} \simeq 31 \ \epsilon \ \ \left[{{\rm m} \over {\rm s}}\right]
\ende
and
\beg
\delta\Omega \simeq 0.058 \ \epsilon \ \ \left[{\rm day}^{-1}\right].
\ende
Both effects, of course,  are negligible for the present-day Sun ($\epsilon \simeq 10^{-5}$) 
but for rapid rotators with (say) $\epsilon \simeq 0.5$ the effects should be 
observable. With (\ref{3}) for solar-type stars it is
\beg
\delta \Omega= 7.9 \cdot 10^{-4} \ {{\rm day}\over \tau_{\rm rot}^2} .
\label{deltaom}
\ende
For $\tau_{\rm rot}=12$ h one finds $\delta\Omega \simeq 0.003$ day$^{-1}$, a value which 
cannot  explain the situation described in Fig. \ref{alpha}. For rotation periods 
exceeding 0.5  day and for the used turbulence model the gravity darkening effect  seems  not to play an important  role. As we shall see 
below,  remarkably  higher values for the equator-pole-difference are possible for modified turbulence parameters but only up  to a factor of 6.

Another  finding concerns the meridional flow. It has to do with the 
idea to interpret the solar butterfly diagram as the result of a meridional 
circulation (equatorwards at the bottom of the convection zone) rather than as 
an intrinsic property of the $\alpha\Omega$-dynamo (Wang \ea 1991, 
Choudhuri \ea 1995, Dikpati \& Charbonneau 1999). 
To this end the eddy diffusivity in the convection zone must be much 
smaller than currently believed. A typical value of a suitable magnetic Prandtl number with which 
meridional flows of amplitude of 10 m/s become magnetohydrodynamically important is 
about 50 (K\"uker \ea 2001). 
\begin{figure}
 \hspace{-0.5cm}
\psfig{figure=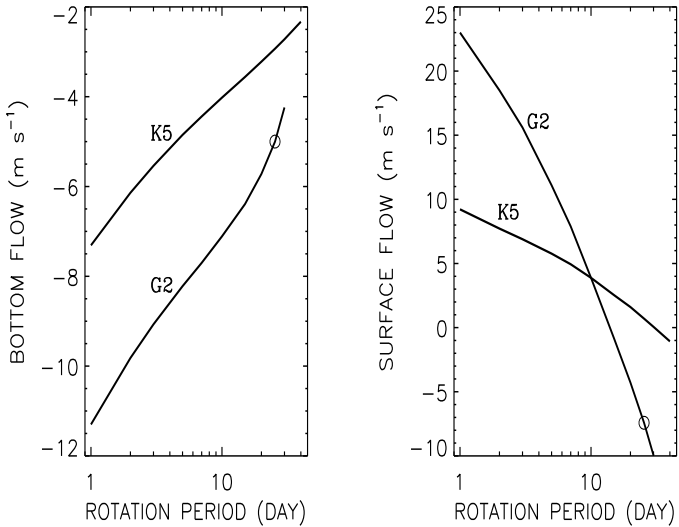,width=9cm,height=7cm}
 \caption{ \label{kr}
The meridional circulation at mid-latitudes (45$^\circ$) at the bottom (LEFT) 
and the top (RIGHT) of the convection zone for both a solar model and a K5 
main-sequence star as a function of rotation period (KR99). The circle represents the present-day Sun. The flow at the bottom is 
equatorwards (as it is the flow at the surface, there are 2 cells in radius)
 }
\end{figure}
The equatorward bottom drifts 
for stars with various rotation rates are given in Fig. \ref{kr}. It varies from 
    5 m/s for the present-day Sun to 11 m/s for a very young solar-type star. The bottom 
    drift for the gravity darkening effect scales as 
\begin{equation} \label{ll}
u_{\rm bot} \simeq -5.8 \cdot \epsilon
\end{equation}
in m/s. It flows opposite to the meridional flow induced by the $\Lambda$-effect, 
without nonuniform heating from below. As a result of the nonuniform heating the 
influence of the meridional flow onto the migration of the stellar toroidal 
field belts is reduced. A new and  interesting question is whether the total meridional flow  can change its 
sign by the influence of nonuniform heating. The amplitude (\ref{ll}), however, 
seems to be too small to overcompensate 10 m/s even for the fastest rotation. 

\begin{table}
 \caption{Meridional flow at top and bottom of the convection zone, equator-pole 
 differences of angular velocity and temperature for $\epsilon=0.1$ (rotation
 period of $\simeq$ 9 h). Poleward circulation is marked with a minus}
\begin{tabular}{|l||c|c|c|c|c|}
\hline
&&&&&\\[-1ex]
$c_\nu$  & 0.1 & 0.1 & 0.3 & 1 & 1 \\
$c_\chi$ & 0.1 & 0.3 & 0.3 & 1 & 0.3\\
Pr       & 1   & 0.3 & 1   & 1 & 3\\[1ex]
\hline
&&&&&\\[-0.5ex]
$u_{\rm top}$ [m/s] \quad\quad  & +10    & +4.1   & +5.3   & +3.1   
& +7.3\\[1.5ex]
$u_{\rm bot}$ [m/s] \quad \quad & -1.3    & -0.46   & -0.81   & -0.58   
 & -1.50\\[1.5ex]
$\delta\Omega$ [day$^{-1}$] \quad\quad & 0.03     & 0.013   & 0.012   
 & 0.005   & 
0.010\\[1.5ex]
$\delta T$ [K] \quad\quad & 2  & 1    & 1    & 0.5    & 1 \\[1ex]
\hline
\end{tabular}
\label{tab1}
\end{table}

We have thus to vary the free eddy diffusion parameters $c_\nu$ and $c_\chi=c_\nu/{\rm Pr}$. The results for a variety of 
parameters $c_\nu$ and Pr are presented in Table \ref{tab1}. In all cases they are written for
 $\epsilon =0.1$, i.e. for a rotation period of about 9 h. There are no strong 
 dependencies of the results on the input 
parameters $c_\nu$ and Pr. We have given in Table \ref{tab1} also 
the heat conductivity parameter $c_\chi=c_\nu/{\rm Pr}$. Indeed, the equator-pole 
difference of the rotation rate has an  inverse trend with $c_\chi$. For 
$c_\chi =1$ we find only 50\% of $\delta\Omega$ resulting for $c_\chi=0.3$. For $c_\chi=0.1$ the factor of the corresponding  $\delta\Omega$ is 6, very close to the observations. 

On the other hand, there is a clear trend of the meridional circulation at the 
bottom of the convection zone with the turbulent Prandtl number. When the Prandtl 
number grows by a factor of 10 the bottom drift increases by a factor of 3.  One can also 
find the reduction of the meridional flow  for decreasing viscosity first described by K\"ohler (1969). 
His models demonstrate  a distinct maximum of 10 m/s for   an eddy viscosity of about 
$5\cdot 10^{13}$ cm$^2$/s.  Density stratification, however, is not included.  Models with density stratification 
require a faster  meridional flow (factor 4 for a density contrast of 200) than 
 models without density stratification (R\"udiger 1989, p. 113).
For an eddy viscosity of $5\cdot 10^{12}$ in a density-stratified convection
zone an equatorward circulation of 8 m/s at the surface produces a equator-pole
difference of the angular velocity of $\delta \Omega \simeq 0.06\  {\rm
day}^{-1}$. The data given in the second column of Table \ref{tab1} confirm this
well-known result (see also Kippenhahn 1963).

In all cases the latitudinal differences of the temperature are unobservably small.
The maximal value for $\epsilon=1$ is only 20 K. This finding is in accordance to 
the finding that the convection zone is able for an effective screening of all 
large-scale temperature differences at the bottom of the convection zone (Stix 
1981).

\end{document}